\documentclass[pra, 12 pt]{revtex4}
\usepackage[english]{babel}
\usepackage{amsmath}
\usepackage{epsfig}

\begin{document}

\title{\bf EXACT LIMITING RELATION BETWEEN THE STRUCTURE FACTORS IN NEUTRON AND X-RAY SCATTERING}
\author{V.B. Bobrov, S.A. Trigger, S.N. Skovorod'ko}
\address{Joint\, Institute\, for\, High\, Temperatures, Russian\, Academy\,
of\, Sciences, 13/19, Izhorskaia Str., Moscow\, 125412, Russia;\\
email:\,satron@mail.ru}

\begin{abstract}
The ratio of the static matter structure factor measured in
experiments on coherent X-ray scattering to the static structure
factor measured in experiments on neutron scattering is
considered. It is shown that this ratio in the long-wavelength
limit is equal to the nucleus charge at arbitrary thermodynamic
parameters of a pure substance in a disordered state.\\

PACS number(s): 31.10.+z, 05.30.-d, 52.27.Gr, 71.10.-w\\

\end{abstract}

\maketitle

According to [1], we proceed from the fact that properties of real
matter are defined by the nature of the Coulomb interaction
reflected in collective behavior of interacting electrons and
nuclei. At the same time, in most applications of statistical
physics, it is conventional to proceed from the effective
Hamiltonian with short-range forces, corresponding to the problem
under consideration. The direct consideration of the\,
"collective"\, Coulomb interaction is mostly performed with
respect to\, "quasi-free"\ electrons and ions (as atoms whose
valence electrons\, "collectivized") or in studying model
single-component plasma (see, e.g., [2]). This situation causes
the terminological difference between the notions\, "Coulomb
system"\, and\, "plasma". Therefore, for further consideration, we
will introduce two notions:

- "simple"\, (or\, "neutral") matter consisting of identical
particles (e.g., of type $a$) interacting with a short-range
potential and

- "pure"\, matter being a quasi-neutral two-component system with
the Coulomb interaction, consisting of electrons (subscript $e$)
and nuclei (subscript $c$),
\begin{eqnarray}
\sum_{\beta\,=\,e,c} e z_\beta n_\beta=0, \label{S1}
\end{eqnarray}
where $n_\beta$  is the average density of the number of particles
of type $\beta$ with charge $e z_\beta$. In most applications,
"simple"\, matter is a system of interacting atoms (or molecules)
with, generally speaking, internal structure. In view of the
above, "simple"\, matter can be considered as a model of\,
"pure"\, matter at certain thermodynamic parameters.
Thermodynamics and statistical physics of\, "simple"\, matter are
well developed (see, e.g., [3]). Moreover, the model of\,
"simple"\, matter makes it possible to successfully enough
describe the data on the matter [4], obtained in experiments on
neutron scattering [5]. The neutron scattering cross section (see,
e.g., [5,6]) is defined by the static nucleus structure factor
$S_{cc}(q)$, directly related to the pair correlation function
$g_{cc}(r)$ of nuclei by the general relation
\begin{eqnarray}
S(q)=1+n\int \exp (-i {\bf q r})\left(g(r)-1\right) d{\bf
r}\label{S2}
\end{eqnarray}
In this case, the functions $S_{cc}(q)$    and  $g_{cc}(r)$ for
nuclei are directly associated with functions $S_{aa}(q)$ and
$g_{aa}(r)$ for atoms. Hence, within the model of "simple" matter,
the well-known limit relation [3-6] is satisfied,
\begin{eqnarray}
\lim_{q\rightarrow 0} \,S_{a\,a}(q)=n_a T K_T, \,\;\;
K_T=-\frac{1}{V}\left(\frac{\partial V}{\partial P}\right)_T,
\label{S3}
\end{eqnarray}
where $n_a$  is the average density of the number of atoms and
$K_T$ is the isothermal compressibility of the system at
temperature $T$ and pressure $P$ in volume $V$. However, in
describing experimental data on coherent X-ray scattering, such a
direct association can no longer be used, since the static
electron structure factor $S_{e\,e}(q)$  is measured in
experiments on X-ray (and light) scattering (see, e.g., [7,8]).
Therefore, when using the model of "simple" matter, there is the
only opportunity of explaining experimental data on light
scattering, i.e., to consider the atom as a nucleus with electrons
localized on it (similarly to the conventional quantum-mechanical
consideration [9]), i.e., as a compound particle. This at first
sight evident statement has not yet been strictly theoretically
justified because of the necessity of the consideration of
electron identity, hence, the atom itself has not yet been
statistically described in the system of interacting electrons and
nuclei (see, e.g., [1,10]).

Nevertheless, within the model of\, "simple"\, matter, it is
accepted to determine the relation between the atom structure
factor $S_{a\,a}(q)$ (strictly speaking, the nucleus structure
factor $S_{c\,c}(q)$ ) and the electron structure factor
$S_{e\,e}(q)$ by putting into consideration (see, e.g., [5,11,12])
the atom form factor $F_{a\,a}(q)$,
\begin{eqnarray}
n_e S_{e\,e}(q)=n_a |F_a(q)|^2 S_{a\,a}(q),\;\; F_a(q)=\int \exp
(-i {\bf q r}) \varrho_e^a(r) d{\bf r},\;\;
 \lim_{q\rightarrow 0} \,F_a (q)=z_c \label{S4}
\end{eqnarray}
where $\varrho_e^a(r)$  is the nonuniform density of electrons
localized on the atom. The last equality in (3) is a consequence
of the electrical neutrality of the atom [9]. Here it is clear
that $n_a=n_c$ . Taking into account (3), representation (4) for
the electron structure factor $S_{e\,e}(q)$ is still used to
theoretically justify the critical opalescence phenomenon (sharp
light scattering enhancement by pure matter in the vicinity of the
critical point, where the isothermal compressibility
$K_T\rightarrow\infty$) (see, e.g., [4,5]).

Nevertheless, it is clear from physical considerations [1,2]
that\, "simple"\, matter at high densities or temperatures
contains\, "quasi-free"\, electrons. Thus, "simple"\, matter
transforms to the\, "plasma"\, state. At the same time, it is
clear that the classification of matter states into\, "neutral"\,
and plasma ones is quite conditional, since a finite number of\,
"quasi-free"\, electrons always exist at nonzero temperatures
(see, e.g., [1,10]). This is equivalent to the conditional
classification of substances into dielectrics and conductors (and
semiconductors), depending on the static conductivity which is
nonzero at nonzero temperature for any matter state [13]. Thus,
representation (4) for the electron structure factor $S_{e\,e}(q)$
is quite conditional even from physical considerations from the
viewpoint of the necessity to consider "quasi-free" electronic
states, which was noted in [14] when considering X-ray scattering
in metals. Thus, from physical considerations, electronic states
in matter can be conditionally classified into two groups, i.e.,
"localized"\, ones forming compound particles with nuclei (ions,
atoms, molecules, etc.) and\, "delocalized"\, ones in which
electrons behave as quasi-free ones [15]. Thus, the average
density of the number of electrons $n_e$ in the system can be
written as the sum of densities of numbers of electrons
$n_e^{loc}$  in localized and $n_e^{deloc}$ in delocalized states:
$n_e=n_e^{loc}+n_e^{deloc}$. The relation between $n_e^{loc}$ and
$n_e^{deloc}$ varies depending on thermodynamic parameters of the
system. Let us further assume that only one-center bound states of
electrons and nuclei exist in the system under consideration,
i.e., there are no molecules and the more complex multinuclear
structures in the system. In this case, only\, "ions"\,
characterized by charge $z_i=n_e^{loc}/n_i=n_e^{loc}/n_c$ can be
compound particles (it is obvious that their density $n_i$
coincides with the density $n_c$ of nuclei). At first sight, such
consideration is identical to the case of fully ionized plasma
(see, e.g., [2]). However, within the above analysis, the\,
"ion"\, charge  $z_i$ is not an integer and continuously varies
from zero to the nucleus charge $z_c$, depending on thermodynamic
parameters of the system. Thus, $z_i$ in this approach is a
statistical quantity in contrast to the traditional consideration,
where quantum mechanical results are applied to ions, atoms,
molecules, etc., within the so-called chemical model of plasma
(see, e.g., [2]). In this sense, "simple"\, matter is an extreme
case of such consideration at $z_i\rightarrow 0$ .

If we further neglect the exchange-correlation interaction between
electrons in localized and delocalized states and between
electrons localized on different\, "ions", we can write the
following relation between the static electron $S_{e\,e}(q)$ and
nucleus $S_{c\,c}(q)$ structure factors (in this case, it is
obvious that $S_{c\,c}(q)=S_{i\,i}(q)$), [16]
\begin{eqnarray}
n_e S_{e\,e}(q)=n_c |F_i(q)+\varrho_e^{deloc}(q)|^2
S_{c\,c}(q),\;\; F_i(q)=\int \exp (-i {\bf q r})
\varrho_e^{loc}(r) d{\bf r},\label{S5}
\end{eqnarray}
\begin{eqnarray}
\varrho_e^{deloc}(q)=\int \exp (-i {\bf q r}) \varrho_e^{deloc}(r)
d{\bf r},\;\; \lim_{q\rightarrow 0}
F_i(q)=z_c-z_i,\;\;\lim_{q\rightarrow 0}\varrho_e^{deloc}(q)=z_i
\label{S6}
\end{eqnarray}
where  $F_i(q)$ is the\, "ion"\, form factor,
$\varrho_e^{deloc}(q)$ is the form factor of\, "delocalized"\,
electronic states (to determine which, the perturbation theory on
the electron-ion interaction should be used), $\varrho_e^{loc}(r)$
is the inhomogeneous electron density of states localized on one\,
"ion"\,,  $\varrho_e^{deloc}(r)$ is the inhomogeneous density of\,
"delocalized"\, electronic states per\, "ion"\, in the static
field of an\, "ion"\, set. From (5), (6), taking into account the
quasineutrality condition (1), it immediately follows that
\begin{eqnarray}
\lim_{q\rightarrow
0}\frac{S_{e\,e}(q)}{S_{c\,c}(q)}=z_c,\label{S7}
\end{eqnarray}
i.e., the ratio of the static structure factor of matter,
determined in experiments on coherent X-ray scattering to the
static structure factor determined in experiments on neutron
scattering is equal to the nucleus charge of a given substance in
the long-wavelength limit. We note that a similar equality takes
place in the model of\, "simple"\, matter, which follows from (1)
and (4). According to the above consideration, the statement (7)
should not depend on thermodynamic parameters of matter. On the
other hand, it is clear that the electron structure factor
$S_{e\,e}(q)$  of matter differs significantly from the nucleus
$S_{c\,c}(q)$ structure factor of matter. Taking into account (5)
and (6), the degree of this difference can be expressed by a
certain function $F(q)$,
\begin{eqnarray}
n_e S_{e\,e}(q)=n_c |F(q)|^2 S_{c\,c}(q).  \label{S8}
\end{eqnarray}
In this regard, it should be noted that not true values of the
electron structure factor $S_{e\,e}(q)$  are given in the
overwhelming majority of papers devoted to experimental X-ray
diffraction studies of disordered matter, but recalculated data on
the structure factors $\tilde S_{e\,e}(q)$, obtained from (8)
using certain approximation for the function $F(q)$  according to
(4) or (5), (6),
\begin{eqnarray}
\tilde S_{e\,e}(q)=\frac{n_e S_{e\,e}(q)}{n_c |F(q)|^2}.
\label{S9}
\end{eqnarray}
with the requirement of the maximum closeness of the functions
$\tilde S_{e\,e}(q)$ and $S_{c\,c}(q)$ (see [12, 16-19] and
references therein). However, the form factor $F(q)$, as the above
consideration showed,  can be calculated only approximately and in
a limited range of parameters. Thus, experimental data on coherent
X-ray scattering after such recalculation are to a large extent
applied to test the adequacy of the model for calculating the form
factor $F(q)$ and, using the latter, to fit the structure factor
$\tilde S_{e\,e}(q)$ to the nucleus structure factor $
S_{c\,c}(q)$ which can be determined independently from
experiments on neutron scattering. In essence, the case in point
is testing the models for the form factor $F(q)$. One of such
reliable enough theoretical models for the form factor $F(q)$ is
exactly its representation in the form of (5), (6). In principle,
based on comparison of experimental data on neutron and X-ray
scattering at identical thermodynamic parameters, such
$F(q)=F^{exp}(q)$ can always be found, which will satisfy the
requirement $\tilde S_{e\,e}(q)=S_{c\,c}(q)$. From (1), (7), (8),
it immediately follows that
\begin{eqnarray}
\lim_{q\rightarrow 0}F(q)=z_c,\label{S10}
\end{eqnarray}
The statements (7) and (10) obtained from physical considerations
can be proved for "pure" matter, based on general limit relations
for the correlation functions of the multicomponent Coulomb
system, obtained in [20] using the of diagram technique of the
perturbation theory (see, e.g., [21]). According to [20], for
static structure factors of "pure" matter as a disordered
two-component Coulomb system, the equalities are valid in the
non-relativistic limit, which represent a direct consequence of
the Coulomb nature of the interparticle interaction,
\begin{eqnarray}
\lim_{q\rightarrow 0} \,S_{c\,c}(q)=n_c T K_T; \;\;
\lim_{q\rightarrow 0}
\,S_{c\,c}(q)=\frac{n_c}{n_e}\,\lim_{q\rightarrow 0}\,
S_{e\,e}(q)= \left(\frac{n_c}{n_e}\right)^{1/2}\lim_{q\rightarrow
0} \,S_{e\,c}(q) \label{S11}
\end{eqnarray}
Taking into account the quasineutrality condition (1), the limit
equalities (7) and (10) immediately follow from (11). Moreover,
relation (11) provides a rigorous theoretical justification to the
critical opalescence phenomenon. Furthermore, relations (11) are
based on the use of the diagram technique of the perturbation
theory [21], whose applicability limits cannot be determined from
general considerations. Thus, the general nature of statements (7)
and (10) can be considered as the possibility of experimental
confirmation of the strict result (11) of the theory of disordered
Coulomb systems, since the diagram technique of the perturbation
theory is, in essence, a unique consistent method of the
theoretical study of quantum systems of interacting particles.
Thus, the main result of this paper is the indication of the
existence of the exact relation valid at any thermodynamic
parameters of disordered "pure" matter,
\begin{eqnarray}
S_{e\,e}(q\rightarrow 0)=z_c S_{c\,c}(q\rightarrow 0) \label{S12}
\end{eqnarray}
This relation can be verified in experiments on X-ray and neutron
scattering, performed at arbitrary (but identical) densities and
temperatures corresponding to the disordered (not crystalline)
state of substances under study. Although it is difficult to
experimentally provide extremely small scattering angles, hence,
wave vectors $q$, the trend toward satisfying equality (12) at
small $q$ should be clearly observed. The importance of the
proposed validation of relation (12) consists in its generality
and universality. Moreover, such a test seems to be extremely
important to confirm the fundamental theoretical approach to the
description of the disordered state of matter as a Coulomb system
of electrons and nuclei (the electron-nuclear model of matter).

This study was supported by the Netherlands Organization for
Scientific Research (NWO), grant no. 047.017.2006.007 and the
Russian Foundation for Basic Research, projects no. 07-02-01464-a
and no. 10-02-90418-Ukr-à.

\end{document}